\documentclass{aa} 
\usepackage{color}
\usepackage{graphicx}
\usepackage{tabularx}

\usepackage{rotating}
\usepackage{txfonts}
\usepackage{comment}
\usepackage{lscape}
\usepackage{threeparttable}
\usepackage{xcolor}
\usepackage{subfigure}
\usepackage{float}
\usepackage{ulem}

\newcommand{\HI}{{\ion{H}{1}}}

\newcommand{\kms}{$\,$km$\,$s$^{-1}$}
\newcommand{\WHz}{$\,$W$\,$Hz$^{-1}$}

\newcommand{\mJybeam}{mJy beam$^{-1}$}
\newcommand{\Jybeam}{Jy beam$^{-1}$}

\newcommand{\msun}{{$M_\odot$}}
\newcommand{\msunyr}{{$M_\odot$ yr$^{-1}$}}

\newcommand{\tspin}{$T_{\rm spin}$}

\newcommand{\coTwo}{{CO(2-1)}}

\newcommand{\alwmath}[1]{\ifmmode#1\else$#1$\fi}

\def\HI{\ion{H}{i}}
\def\slHI{{\sffamily H\,\tiny I}}

\def\emph#1{{\sl #1}}
\newcommand{\ltsima} {$\; \buildrel < \over \sim \;$}
\newcommand{\gtsima} {$\; \buildrel > \over \sim \;$}
\newcommand{\lta} {\lower.5ex\hbox{\ltsima}}
\newcommand{\gta} {\lower.5ex\hbox{\gtsima}}

\newcommand{\p}[1]{$^{-#1}$}

\newcommand{\apx}{$\sim$}
\usepackage{txfonts}
\begin{document} 

   \title{Cold gas in the heart of Perseus\,A} 

\authorrunning{Morganti et al.}
\titlerunning{Cold gas in the heart of Perseus\,A}
\author{Raffaella Morganti\inst{1,2}, Suma Murthy\inst{3}, Tom Oosterloo\inst{1,2}, Jay Blanchard\inst{4}, Claire Cook\inst{5}, Zsolt Paragi\inst{3}, \\ Monica Orienti\inst{6}, Hiroshi Nagai\inst{7,8}, Robert Schulz\inst{1}}

\institute{
ASTRON, the Netherlands Institute for Radio Astronomy, Oude Hoogeveensedijk 4, 7991 PD, Dwingeloo, The Netherlands. 
\and
Kapteyn Astronomical Institute, University of Groningen, Postbus 800,
9700 AV Groningen, The Netherlands
\and
Joint Institute for VLBI ERIC, Oude Hoogeveensedijk 4, 7991 PD Dwingeloo, The Netherlands.
\and
National Radio Astronomy Observatory, P.O. Box O, 1003 Lopezville Rd., Socorro, NM 87801
\and
Department of Physics \& Astronomy, University of Kansas, 1251 Wescoe Dr., Lawrence, KS 66045, USA
\and INAF - Istituto di Radioastronomia, Via P. Gobetti 101, I-40129 Bologna, Italy
\and
National Astronomical Observatory of Japan, 2-21-1 Osawa, Mitaka, Tokyo 181-8588
\and 
The Graduate University for Advanced Studies, SOKENDAI, Osawa 2-21-1, Mitaka, Tokyo
181-8588}
 \abstract
{We present new Karl G.\ Jansky Very Large Array (JVLA) and Very Long Baseline Array (VLBA) observations that traced the \HI\ in the central region of 3C~84 (Perseus~A). This radio  source is hosted by the bright cluster galaxy NGC~1275 in the centre of the iconic Perseus cluster.  With the JVLA, we detected very broad (FWHM \apx 500 \kms) \HI\ absorption at
 arcsecond resolution (\apx 300 pc)  that is centred at the systemic velocity of NGC 1275 against the bright  radio continuum, suggesting that the detected gas is very close to the supermassive black hole (SMBH). However, we did not detect any  absorption in the higher-resolution VLBA data against the parsec-scale radio core and  jet. 
Based on a comparison of the properties of the \HI\ absorption with those of the molecular circum-nuclear disc (CND) known to be present in NGC 1275, we argue that the \HI\ seen in absorption arises from  \HI\ in this fast-rotating CND, and that neutral atomic hydrogen is present  as close as \apx 20 pc from the SMBH. The radio continuum providing the background for absorption arises from  non-thermal synchrotron emission from the star formation activity in the CND, whose presence has been reported by earlier  VLBA studies. We did not detect any signature that the  \HI\ gas is affected by an interaction with the radio jet. Thus, at this stage of the evolution of the source, the impact of the radio jet on the gas in NGC 1275 mainly creates  cavities on much larger galaxy scales. This   prevents the circum-galactic gas from cooling, and it does  not produce gas outflows. Overall, the properties of the CND in Perseus~A present a very similar case to that of Mrk~231, suggesting that, unlike often assumed, \HI\ absorption can arise against the radio emission from star formation in a CND and is perhaps common  in radio AGN. With the JVLA, we serendipitously detected  a new,  faint absorbing system that is redshifted by \apx 2660 \kms, in addition to the already known high-velocity
absorption system that is redshifted 2850 \kms\ with respect to NGC 1275. We identify this new system as gas that is stripped from a foreground galaxy falling into the Perseus cluster. This new absorption remains undetected with the VLBA. }
   \keywords{galaxies: active - radio lines: galaxies - galaxies: individual: 3C~84 (Perseus A)}
   \maketitle  


\section{Introduction}
\label{sec:Introduction}

Even in the harsh central regions of galaxies hosting an active galactic nucleus (AGN), observations of the cold gas (in the form of atomic neutral hydrogen and cold molecular gas) allow us to trace a variety of phenomena (\citealt{Morganti18,Veilleux20,Combes22}).
Not only can cold gas  be present in circum-nuclear disc (CND) structures  with sizes of a few hundred parsec, it also traces infalling gas associated with the feeding of the supermassive black hole (SMBH) \citep{Tremblay16,Maccagni18,Rose19,Combes22} and gas that is outflowing under the effect of winds and radiation (\citealt{Brusa18,Veilleux20} and refs therein) or under the effect of the mechanical impact from radio jets (\citealt{Oosterloo17,Murthy22,Girdhar22,Audibert23}).
One of the advantages of tracing this cold gas is the possibility of observing it at the very high spatial resolution allowed by radio telescopes. The gas can thus be traced in the circum-nuclear region, very close to the SMBH and in the region that is co-spatial with the inner parts of the jets.

The central regions of galaxy clusters, and in particular, cool-core galaxy clusters,  are examples of the complicated symbiosis between AGN and the gaseous intra-cluster medium that includes the cold gas. This makes them ideal targets for the study of all the phenomena listed above. The vast majority of these clusters host a radio galaxy in their centre. The radio plasma emitted by these galaxies   in the form of jets and lobes is thought to play a key role in preventing the catastrophic cooling of the hot cluster medium. Nevertheless, many of these clusters do contain cold gas. 
Cold molecular gas  mostly traced by CO lines has been observed in many cases (see e.g.\ \citealt{Russell19,Tremblay16,Rose23}). ALMA observations have shown that the  molecular gas in cool-core clusters is often distributed in dynamically unstable filamentary structures associated with X-ray bubbles inflated by radio lobes and resulting from  the cooling induced by this interaction (see \citealt{Russell19,Lim08}). So far, only a minority of cases appear to have  CO in circum-nuclear discs, such as Hydra~A and 3C 84 (\citealt{Rose19,Nagai19} respectively), with the latter being the target of our study.

Atomic neutral hydrogen (\HI) is also found to be present in  radio galaxies hosted at the centre of these clusters via the detection of the 21 cm line in absorption (see \citealt{Morganti18} for an overview). The studies of this particular topic have not been extensive so far, however. A search for \HI\ absorption in a small sample of bright central galaxies (BCGs) in clusters by \cite{Hogan14} found a few detections in which the absorption was either at the systemic velocity or mildly redshifted. A number of studies focusing on individual radio sources in clusters were conducted, for example PKS~2322--123 in Abell~2597 (\citealt{ODea94,Taylor99}), 3C~84 in the Perseus cluster (\citealt{Sijbring89,Jaffe90}),  4C~26.42 in Abell~1795 (\citealt{Bemmel12}), 4C~35.06 in Abell~407 (\citealt{Shulevski15}), Hydra~A \citep{Taylor96}, Cygnus~A \citep{Conway95,Struve10c}, and the Norma cluster \citep{Saraf23}.
Thus, radio galaxies in cool-core clusters are interesting objects in which several sometimes competing processes can be studied using the combination of CO in emission and \HI\ absorption (see e.g.\  \citealt{Tremblay16}).

Here we present new observations that trace the \HI\ in absorption in 3C 84  (Perseus A). This is one of the most famous radio galaxies and is located in the centre of the iconic gas-rich cool-core Perseus cluster. The observations cover  parsec to kiloparsec scales. We connect the results with the properties of the molecular gas and of the radio continuum. As described in Sect.\  \ref{sec:3C84-overview},  molecular gas in this source has been detected in the form of filaments and a CND. Interaction between the jet and surrounding gas clouds has been suggested based on the VLBI continuum structure of the radio jet \citep{Nagai17,Kino18,Kino21}. With the \HI\ absorption presented here, we investigate the interplay between these structures. 

This paper is structured in the following way. In Sect.\ \ref{sec:3C84-overview} we briefly summarise some of the relevant results from the extensive literature of this well-known source to set the context for our study. In Sect.\  \ref{sec:Observations} we describe  new observations we obtained with the Karl G.\ Jansky Very Large Array (JVLA) and with the Very Long Baseline Array (VLBA). In Sect.\ \ref{sec:Results} we present the results on the radio continuum and the \HI\ on arcsec and milliarcsec scales. In Sect.\  \ref{sec:Discussion} we combine these results and connect them with those  obtained  on the molecular gas from ALMA observations (\citealt{Nagai19,Oosterloo23}) in order to derive a picture of the distribution of \HI. We conclude by discussing the implications of our results.

\begin{figure*}
  \centering
\includegraphics[width=18cm]{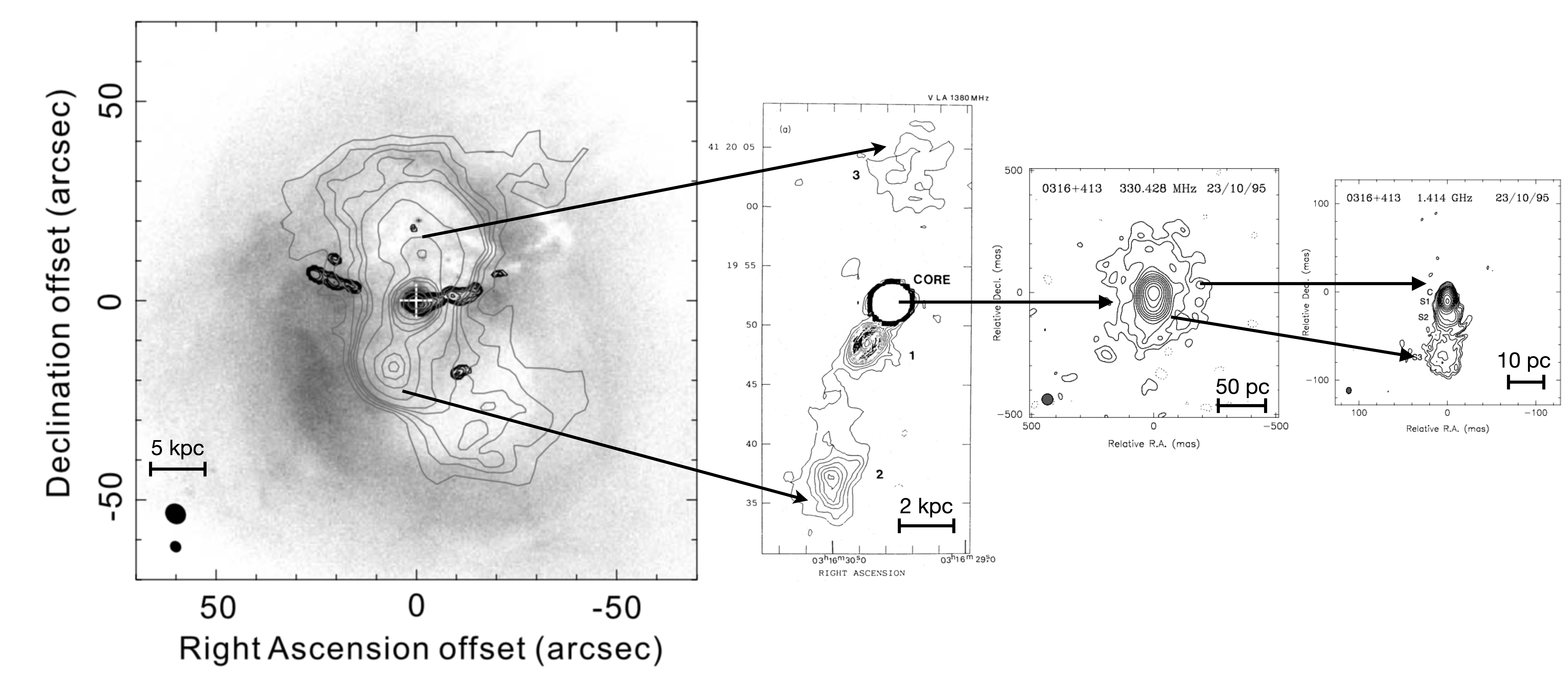}
 \caption{Images of 3C~84 system from the kiloparsec to parsec scales relevant for this study. {\sl Left:} Continuum map at 90cm (\citealt{Pedlar90};  angular resolution of \apx 5$''\simeq 1.8$ kpc) in grey contours superimposed on the map of molecular gas (\citealt{Lim08};  resolution 3$''$  $\equiv 1.1 $ kpc, black contours), with the X-ray image from \cite{Fabian11} in the background. {\sl Second panel from the left:} Image at  \apx 1 arcsec $\equiv 360$ pc  resolution at 1380 MHz from \cite{Pedlar90} showing the continuum structure against which the \HI\ is seen in absorption. {\sl \textup{The last two panels on the right}} were obtained using the VLBA at 330 MHz and 1.4 GHz by \cite{Silver98} and show, in addition to the core-jet structure (right), a diffuse, extended structure (which they call mini-halo) on larger scales (\apx 200 pc  $\equiv$ 0\farcs5  radius).}
\label{fig:ContinuumLiterature}
\end{figure*}

\section{3C~84 and its host galaxy}
\label{sec:3C84-overview}

The radio galaxy 3C~84 is  associated with the BCG NGC~1275 in the Perseus cluster\footnote{
For the cosmology adopted in this paper, we assumed a flat Universe with the following parameters: $H_{\circ} = 70$ \kms\ Mpc$^{-1}$, $\Omega_\Lambda = 0.7,$ and $\Omega_{\rm M} = 0.3$. For the redshift of 3C 84 of $z = 0.01756$ (systemic velocity $V_{\rm sys} = 5264$ \kms; \citealt{Huchra99}), this gives a luminosity distance of 76.2 Mpc and an angular scale distance of 73.6 Mpc so that 1 mas = 0.357 pc.}.
It has an intermediate radio power ($P_{\rm 1.4~GHz} \sim 10^{25}$ \WHz; \citealt{Fabian11,Pedlar90}), and the radio continuum emission covers a large set of  scales, each showing interesting properties. The morphology of the radio continuum emission is shown (in contours) in Fig.\ \ref{fig:ContinuumLiterature}, ranging from kiloparsec to parsec scales and highlighting the features relevant for the present study. 
The image on the left shows the large-scale structure (about 80 arcsec corresponding to about 29 kpc), which appears relatively symmetric. This symmetry is lost at  arcsecond resolution (second panel from the left, from \citealt{Pedlar90}), where the southern jet is clearly brighter. The bright core  also shows a complex structure, as is visible in the two right panels of Fig.\ \ref{fig:ContinuumLiterature}. A combination of diffuse emission (detected by \citealt{Silver98} and called a mini-halo) and a bright jet to the south are present. These structures on the sub-arcsecond scale are particularly relevant for our study.

The inner region of the 3C~84 radio jet has been studied in detail at very high spatial resolution \citep{Giovannini18}. Monitoring of the parsec-scale region has revealed the continuous emergence of new components in the jet    \citep{Savolainen21,Nagai10,Suzuki12}. A sharp bending of the jet has also been observed \citep{Nagai17,Kino18,Kino21}. \cite{Nagai17} reported an abrupt change in the position of the hotspot of the innermost jet, as well as an increase in polarised emission on parsec scales that they interpreted  as showing that the jet interacts with the surrounding inhomogeneous and clumpy medium. However, the change in the position angle of the jet was also explained by \cite{Suzuki12} as caused by the difference in the path followed by newly ejected components from the path of the previously ejected components.  
If we were able to identify whether gas is present with disturbed kinematics that trace this interaction, we would be able to distinguish between these two scenarios and shed light on the inner structure of this complex radio galaxy. This is one of the goals of our study.

As expected in a cool-core cluster, its central region is gas rich.  3C~84 is indeed known to be embedded in ionised and molecular gas as well as \HI. Ionised gas is seen in  the form of a complicated web of large-scale H$\alpha$ filaments out to 50 kpc (\citealt{Conselice01,Fabian08}) and is traced in the central regions by the detection of free-free absorption (\citealt{Walker00}) and through [FeII] and Br$\gamma$ lines from the CND (\citealt{Scharwachter13,Riffel20}). Molecular gas is also seen in large-scale filaments out to large radii \citep{Salome06}, as well as on much smaller arcsecond and sub-arcsecond scales in the central regions \citep{Lim08,Scharwachter13,Nagai19}. \cite{Lim08} used the the Submillimeter Array (SMA) and detected filaments in \coTwo\ on kiloparsec scales, with kinematics consistent with radial infalling motion and without a clear signature of  rotation. Interestingly, at the higher spatial resolution (0\farcs1 $\times $ 0\farcs05 $\equiv 36 \times 18$ pc) obtainable with ALMA, \cite{Nagai19} detected \coTwo, HCN(3-2) and HCO$^+$(3-2) emission distributed in a CND-like structure  with a diameter of 100 pc (0\farcs3), with a mass of cold H$_2$ of $4\times10^8$ \msun. This CND is also detected in warm H$_2$ \citep{Wilman05,Scharwachter13} and  could provide a reservoir of gas to feed the SMBH. This CND consists of a fast Keplerian rotating inner part with a diameter of about 50 pc  and position angle 68$^\circ$ that is surrounded by a  warped disc-like structure of gas (about 100 pc in diameter), the kinematics of which is less regular because large filaments of gas accrete onto the CND (see \citealt{Oosterloo23} for more details).

\begin{figure}
  \centering
\includegraphics[width=\linewidth]{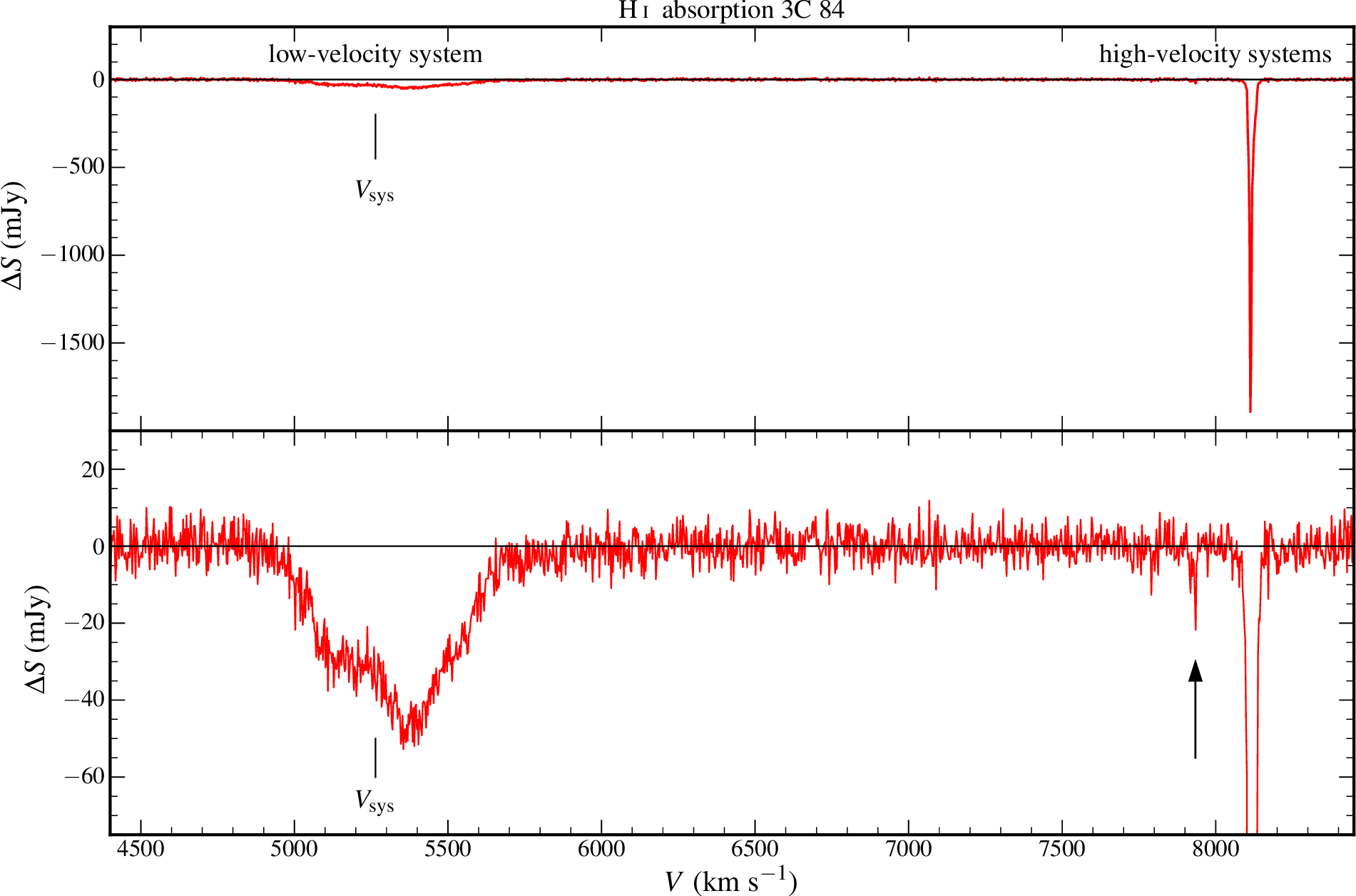}
 \caption{Full  \HI\ profile obtained with the JVLA showing the various absorption systems. We focus on the low-velocity system, centred on the systemic velocity of NGC~1275. The high-velocity absorption is briefly discussed in the appendix. {\it Top:} Spectrum using the full intensity range. {\it Bottom:} Same as above, but with a different vertical scaling. The newly discovered component is indicated with an arrow.}
\label{fig:WSRT}
\end{figure}

Unsurprisingly, \HI\ has also been detected in the form of absorption against 3C~84. The presence of this \HI\ has been known since the low-resolution (kiloparsec scale) observations  by \cite{Crane82}, \cite{Sijbring89}, and \cite{Jaffe90}. They found two absorbing systems: a narrow, highly redshifted system at \apx 8100 \kms\ (the so-called high-velocity system), and a broad, so-called low-velocity, system at the systemic velocity. This is illustrated in the full \HI\ profile in Fig.\ \ref{fig:WSRT} from the JVLA observations presented in this paper.  

The high-velocity absorption is a very high opacity system that has been studied in great detail at high spatial resolution with the VLBA by \cite{Momjian02}. This component is thought to be associated with a foreground cluster galaxy  that is falling towards NGC 1275 but  is still \apx 100 kpc distant from the AGN host galaxy (\citealt{Gillmon04,Sanders07}).

The broader low-velocity system, on the other hand, has not been studied as extensively because the strong continuum emission of 3C~84 requires high-quality bandpass calibration.
Given the complexity of the radio continuum of 3C~84, observations spanning a range of spatial resolutions are required to give us a better picture of the location and conditions of the \HI\  in the heart of Perseus A. This is the focus of our study.  We used 15$^{\prime\prime}$ scales (i.e.\ \apx 5 kpc) with the observations performed during the commissioning of the Multi-Frequency Front End of the Westerbork Synthesis Radio Telescope (WSRT) in April 2001 \citep{Morganti20},  at  arcsecond scales (i.e.\ a few hundred parsecc) with the JVLA and  at milliarcsecond scales (i.e.\ a few dozen parsec) using the VLBA.  

\begin{figure}
  \centering
\includegraphics[width=\linewidth]{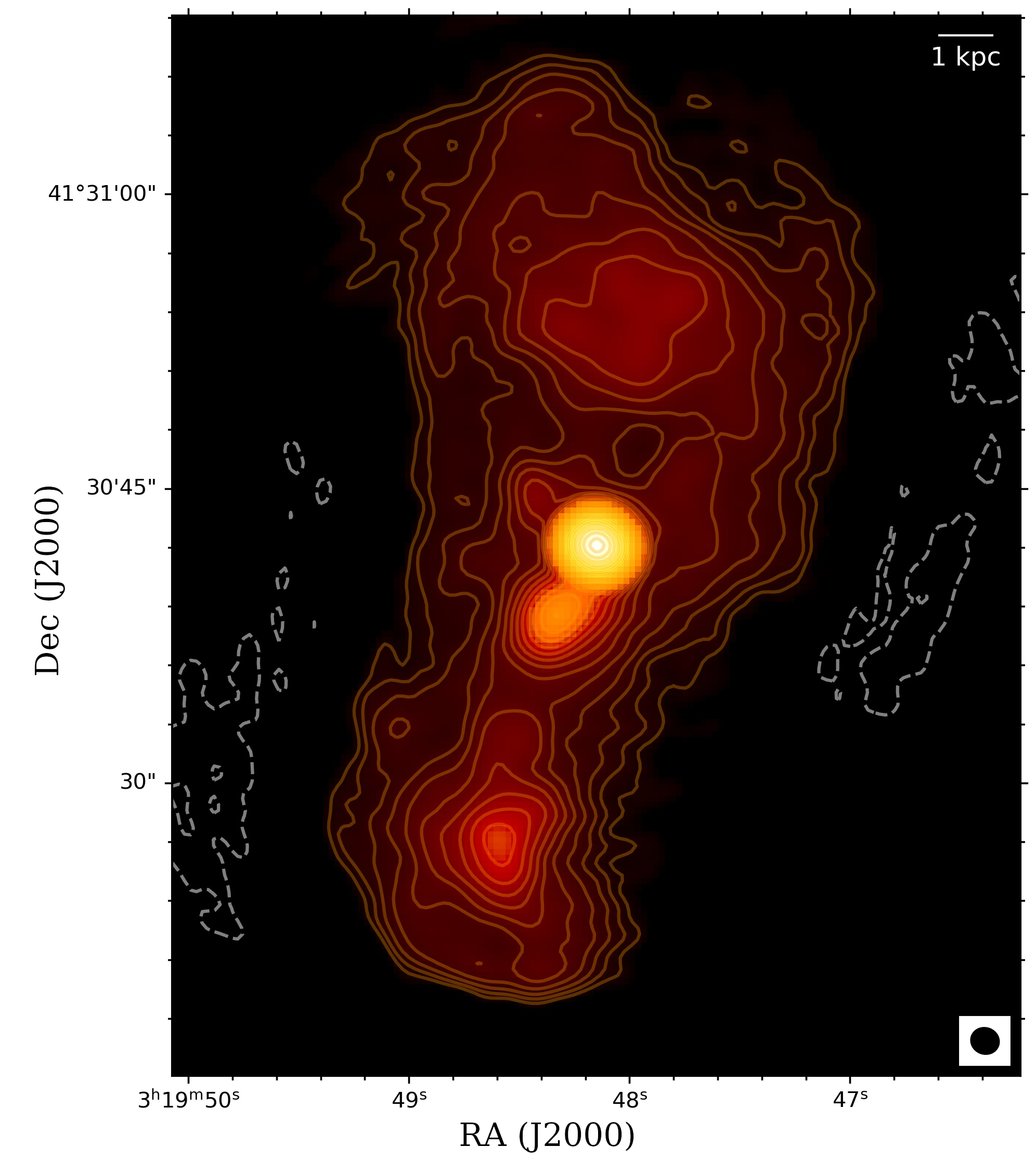}
 \caption{Natural weighted JVLA continuum image. The image has an RMS noise of \apx 0.4 mJy beam\p{1} and a beam size of $1.64'' \times 1.52''$ $\equiv$ 585 pc $\times$ 543 pc  with a position angle of 68\fdg9, shown in the bottom right corner. The contours start at 3 $\sigma$ and increase with factors of $\sqrt{2}$; the 3$\sigma$ negative contours are shown in grey. The image shows a range of structures starting from the bright unresolved core, which has a peak flux density of 14.3 Jy beam\p{1}, the southern jet, diffuse lobes, and also diffuse radio emission connecting all these structures.}
\label{fig:ContinuumVLA}
\end{figure}


\begin{table*}
\caption{Observations and imaging results}
\centering
\begin{tabular}{ccccccccccc}
\hline\hline
Telescope & $\nu_{\rm{obs}}$ & on-source & BW  & $\Delta V$ & Weighting & beam size &PA & RMS$_{\rm{map}}$ & RMS$_{\rm{cube}}$ \\
  & (GHz) & (hours) & (MHz) &(km s$^{-1}$) & & ($''\times''$) &($^\circ$) & (mJy beam\p{1}) & (mJy beam\p{1}) \\
(1) & (2) & (3) & (4) & (5) &(6) & (7) & (8) & (9) & (10)\\ 
\hline
JVLA  & 1.3937 & 9.0 &  32 & 2.7 & Natural & 1.6 $\times$ 1.5 & $\phantom{-}$68.9  & 0.34 & 0.5 \\
      &        &     &    &      & {Robust $=-1$} & 1.1  $\times$ 1.0  & $\phantom{-}$37.0  & 0.45  & 0.7\\
VLBA & 1.3956  & 6.6 & 16 & 33.6 & Natural & 0.009 $\times$ 0.007 & $-$28.5 & 2.7 & 1.0 \\ \hline
\end{tabular}
\begin{tablenotes}
\item Observation details: The columns list  (1) the telescope, (2) the central frequency, (3) the observing time on source, (4) the total bandwidth (5) the velocity resolution, (6) the weighting, (7) the beam size, (8) the beam position angle, (9)  the RMS noise in the continuum map, and (10) the RMS noise in the spectral cube at the velocity resolution listed in column (5).
\end{tablenotes}
\label{tab:obs_details}
\end{table*}


\section{Observations}
\label{sec:Observations}

\subsection{JVLA observations} 
\label{sec:VLAobs}

These observations were carried out with the JVLA A array in August 2019 (Project ID: VLA/2018-06-101) over three epochs of four hours each. We spent a total of nine hours on the target with interleaved scans on 3C~147 every 40 minutes that we used for flux, phase, and bandpass calibration. Since at the resolution of our observation the target is very bright with a peak continuum flux density of 14.3 Jy beam$^{-1}$, we spent a total of three hours on the calibrator, which has a total flux of about 22 Jy, in order to reach a good S/N for the bandpass calibration. The observational setup consisted of nine spectral windows, roughly covering the frequency range of 1 -- 2 GHz. Eight of these  were dedicated to continuum imaging with a bandwidth of 128 MHz subdivided into 64 frequency channels for each window. The remaining spectral window was used to study the \HI\ and had a bandwidth of 32 MHz, centred at 1393.7 MHz (i.e.\ close to the absorption frequency of the low-velocity system of 3C~84), divided into 2560 frequency channels, giving a velocity resolution of 2.7 \kms. The eight continuum spectral windows  were found to be strongly affected by radio frequency interference (RFI), and much of the data could not be used. Therefore, all our results are based on   the spectral-line window alone, upon which  the impact of RFI was minimal.

The data reduction was  made using the classic Astronomical Image Processing Software \citep[AIPS;][]{Greisen03}. We used 3C~147 to estimate the antenna-dependent gain solutions. Since 3C~147 is slightly resolved with the A array, we imaged the calibrator and used the model for bandpass calibration. Since 3C~147 is only 1.5 times brighter than the target, we followed the strategy used by
\citet{Oosterloo05} for the bandpass calibration of a very bright source to reduce the impact of the propagation of bandpass errors into the data cube. This entailed  Hanning-smoothing the bandpass solutions by eight channels and then applying these solutions to the unsmoothed 3C~84 data. Next, we carried out self-calibration using only the line-free channels and performed continuum subtraction and de-redshifting  of the $uv$ data to the systemic velocity of 3C~84 using $z = 0.01756$. A detailed description of these steps can be found in \citet{Murthy21}.

\begin{figure}
  \centering
\includegraphics[width=\linewidth]{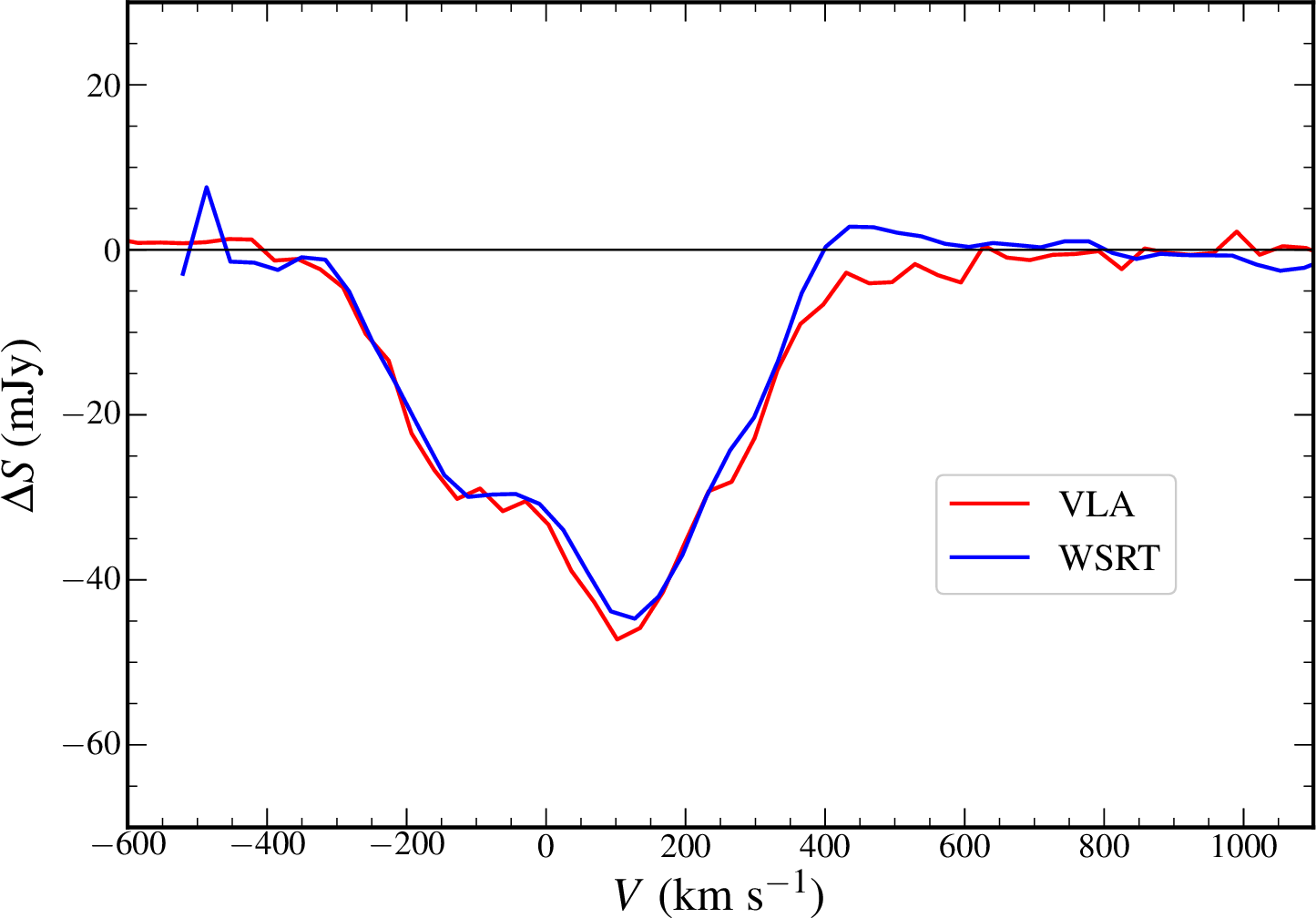}
 \caption{Comparison between the \HI\ spectra  of the low-velocity absorption system obtained with the JVLA and the WSRT. The WSRT spectrum has a velocity resolution of \apx 34 \kms\ , and the JVLA spectrum, originally at a velocity resolution of 2.7 \kms, has been binned to roughly the WSRT channel resolution, followed by a Hanning smoothing. It thus has the same velocity resolution as the WSRT spectrum. Despite the difference in spatial resolution between the two observations (\apx 5 kpc for the WSRT and \apx 350 pc for the JVLA), the strong similarity between the two profiles indicates that the \HI\ producing the absorption and/or the background continuum is located on  sub-kiloparsec scales from the centre (see text for details). Velocities are relative to systemic.}
\label{fig:Comparison}
\end{figure}

We made the continuum images and data cubes at two spatial resolutions: (a) one cube using  natural weighting to recover the diffuse continuum emission and  obtain the highest sensitivity. We made this cube at the raw resolution of 2.7 \kms  . Then, we used (b) a cube using robust = --1 to obtain a better resolution. This latter cube was made at a velocity resolution of 21.6 \kms. The continuum, the spectral beam sizes, and the RMS noise levels we achieved are listed in Table \ref{tab:obs_details}. The naturally weighted continuum image is shown in Fig.\ \ref{fig:ContinuumVLA}, and the integrated spectrum obtained from the naturally weighted cube is shown in Fig.~\ref{fig:Comparison}. Although the sensitivity of the higher-resolution cube is lower, the \HI\ absorption is slightly resolved at this spatial scale, and we use this to compare the kinematics of \HI\ with that of molecular gas in Sec.\ \ref{sec:Properties}.

\subsection{VLBA observations} 
\label{sec:VLBAobs}

The VLBA observations were carried out in March 2019 (Project ID: VLBA/19A-043). No fringes were found to the Pie Town station, and all the baselines to the Mauna Kea station produced strong ripples in the final data cube and thus were  flagged. We observed both left and right circular polarisation, but we did not produce cross-polarisation products as we were only interested in the total intensity. We used a 32 MHz bandwidth, subdivided into 1024 channels, giving a velocity resolution of 6.7 \kms,  centred at the redshifted frequency of the broad \HI\ component. The total observation time was 12 hours. Eight hours were spent on 3C~84, and the remaining time  was spent on interleaved scans on calibrator sources. We used 3C~454.3 and DA~193 as fringe finders and B0307+380 (J0310+3814)  as the phase-reference calibrator. 

The data reduction was made following the standard VLBI data reduction procedure using classic  AIPS and DIFMAP \citep{Shepherd97}. In AIPS we first flagged the data with low weights, corrected for Earth-orientation parameters, applied the ionospheric corrections, corrected for sampler threshold errors from the correlator, and estimated the sampler and gain corrections using the standard AIPS tasks. 
Although we observed various calibrators, precision astrometry was not a main goal of the project, and therefore, we solved for residual fringe rate, delay, and phase solutions directly using 3C84 data (i.e.\ fringe-fitting was made on the target source).
 
We used 3C~454.3 for the  bandpass calibration. Since 3C~84 is brighter than 3C~454.3 and the latter is also extended at VLBI scales, we again used the same technique as with the JVLA data mentioned in Sec.\ \ref{sec:VLAobs} to estimate the bandpass solutions. The calibrated visibilities were then exported to DIFMAP for a few rounds of self-calibration and imaging. This is a powerful and quick way for iterative imaging and self-calibration cycles with additional flagging of data that are poorly (and irreversibly) calibrated or had various issues (instrumental or due to RFI). For the initial steps, we started with uniform weighting, followed by a weighting scheme that balanced  resolution and sensitivity ($uvw = 2, 0$ and $uvw = 2, -1$ in DIFMAP, respectively). For the first several rounds, only the phases were calibrated, initially with a 30-minute solution interval, and gradually going down to 3 minutes. The remaining calibration errors after applying a priori gains (from measured $T_{\rm sys}$ and gains) and the loss of Pie Town meant that it was challenging to recover the most extended emission. 
Before amplitude and phase self-calibration on shorter timescales, finally down to 1 minute, we recovered some of the extended structure by fitting large circular Gaussian components to the residual data, and we continued cleaning with $uvw = 0, -1$ and finally $uvw = 0, -2$ (natural weighting). While this process helped us identify sections of bad data to flag and resulted in a good calibration, the image is still limited by the dynamic range. Some spurious cleaning errors were noted as well. Therefore, we imported the calibrated data to AIPS, which was used to make the final continuum image and the cube. The subsequent steps in the making of the cube are the same as for the JVLA data in Sec.\ \ref{sec:VLAobs}. The final parameters of the images and \HI\ cube are listed in Table \ref{tab:obs_details}.

\section{Results}
\label{sec:Results}

\subsection{Arcsecond scale} 

The JVLA continuum image is shown in Fig.\ \ref{fig:ContinuumVLA}. We detect a range of diffuse structures in addition to the core and jets. Consistent with \cite{Pedlar90}, we detect the bright central component with a peak brightness of $14.3 \pm 0.7$ Jy beam\p{1} and a bright southern jet and lobe extending to $\sim$16$''$ (5.6 kpc) from the core. We do not detect any counter-jet to the north. However, we detect a faint continuum component with a total flux density of $\sim$25 mJy  about 4$''$ north-east from the core. Further to the north, we detect a lobe at \apx $11''$ with a flux density of  \apx 480 mJy. We also detect diffuse radio emission that connects all these components.

The \HI\ spectrum against the bright radio core as obtained from our JVLA data over the entire observing band is shown in Fig.\ \ref{fig:WSRT}. Because the observing band is so wide, we detect the broad \HI\ absorption at the systemic velocity of NGC 1275 as well as the well-known high-velocity absorber redshifted by $\sim$2850 \kms\ with respect to the systemic velocity of NGC 1275. Additionally, we  detect a hitherto unknown absorption feature redshifted by \apx 2660 \kms\ from the systemic velocity of NGC 1275. These two high-velocity absorbers are briefly  discussed in the appendix. All three absorption features are only detected  against the  central core,  and we do not detect absorption against any other continuum components. 

Our focus is on the low-velocity absorber, which we show in detail in   Fig.\  \ref{fig:Comparison}. This absorption feature is particularly broad, with a FWHM $\sim$500 \kms\ , and the profile is relatively symmetric about the systemic velocity. We also show an \HI\ spectrum that we obtained  during test observations  with the WSRT in April 2001 (\citealt{Morganti20}). The spatial resolution of these observations is 15$^{\prime\prime}$. The two profiles are very similar, which  confirms that all \HI\  absorption comes from the central arcsecond region. If the absorption is assumed to arise against the core of 3C~84, we obtain a peak optical depth of $\tau = -\ln(1+\Delta S/c_{\rm f} S_{\rm c})=0.004$ for a peak absorption of 40 mJy beam\p{1} , where $\Delta S$ is the depth of the absorption line, $S_{\rm c}$ is the continuum flux density, and $c_{\rm f}$ is the covering factor, which is assumed to be equal 1. This is lower than the typical optical depths of 0.01 -- 0.02  found  in  radio galaxies \citep{Morganti18}. From integrating the VLA spectrum, we find $\int \tau\ dv = 1.34 \pm 0.07$ \kms\ . The standard formula for estimating the \HI\ column density 
\[
N_{\HI} = 1.82 \times 10^{18} \ T_{\rm spin} \int \tau\,dv \ \ \ \  {\rm cm}^{-2}, 
\]
where \tspin\ is the \HI\ spin temperature, gives a  \HI\ column density of $2.4 \pm 0.1 \times 10^{20}$ cm\p{2} for \tspin\ = 100 K. However, we show in Sec.\ \ref{sec:Origin} that the absorption does not arise against the bright inner core, but against a much fainter extended continuum component, implying  much higher  column densities.

\subsection{Milliarcsecond scale} 

\begin{figure}
  \centering
\includegraphics[width=\linewidth]{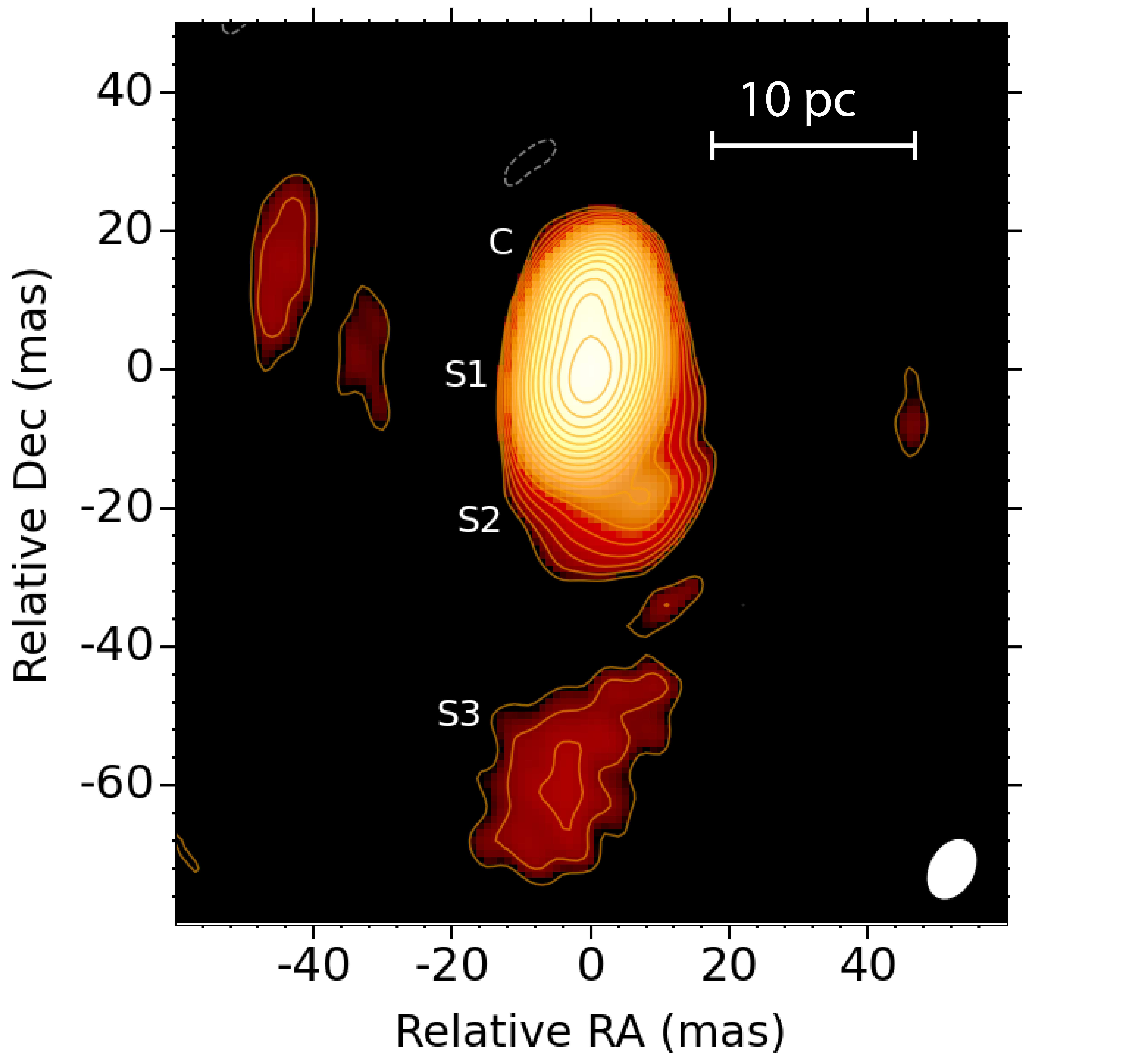}
 \caption{ Naturally weighted VLBA continuum map. The map has an RMS noise of 2.1 mJy beam\p{1} and a beam of 9.4 mas $\times$ 6.7 mas (3.4 pc $\times$ 2.4 pc), with a position angle of --28\fdg5. The contours start at 3 $\sigma$ and increase with a factor of $\sqrt{2}$. The 3$\sigma$ negative contours are shown in grey. We detect the core (C), the southern bright jet (S1), its diffuse extension (S2), and the diffuse lobe farther south of the core (S3). We measure a peak flux density of 3.5 Jy beam\p{1} at S1 and an integrated flux of 9.14 Jy (which includes C+S1+S2+S3).
 }
\label{fig:ContinuumVLBA}
\end{figure}

The VLBA continuum image is shown in Fig.\ \ref{fig:ContinuumVLBA}. Following the designations of \cite{Silver98}, we detect the core (C), the bright southern jet (S1), a diffuse extension from the jet (S2), and the diffuse lobe (S3) about \apx 80 mas (285 pc) farther south from the core (see Fig.\ \ref{fig:ContinuumLiterature}, right). The peak emission is 3.5 Jy beam\p{1}, coincident with S1, while the core at the resolution of our observations is  \apx 1 Jy. We measure an integrated flux of $9.1 \pm 0.9$ Jy. \cite{Silver98} reported a total flux of 22.1 Jy at 1.3 GHz, \apx 2.4 times brighter than our measurement. 
The flux of 3C~84 has varied strongly in recent years (see \citealt{Paraschos23}, although their data cover higher frequencies than ours). This could explain the difference between the flux we measured and that of previous VLBA observations at 1.4~GHz \citep{Silver98,Momjian02}. 

However,  the main result from the VLBA observations is that no trace of the \HI\ absorption feature at the systemic velocity  seen with the JVLA is detected. When the VLBA cube is smoothed to a velocity resolution of 50 \kms, the noise level  is  0.4 mJy beam\p{1}. Much of the absorption in the JVLA spectrum is at the level of 30 --  40 mJy, therefore we should have detected this absorption feature at  high significance if it occurred against the bright inner continuum.  
Narrow features at the 3$\sigma$ level are tentatively seen, but because they are typically very narrow, none was considered a convincing detection. In any case, even if they were real detections, they would be too faint and narrow to account for even a small fraction of the JVLA \HI\ absorption.  Using the noise level, we can derive the upper limit in optical depth and column density of the \HI\ on pc scales. For an rms noise level of  1 \mJybeam\ and a peak flux density of  3.5 \Jybeam\  at S1 and 0.8   \Jybeam\ at the core position, we derive a 3$\sigma$ upper limit of $\tau < 0.0006$  against S1 and $\tau < 0.0026$ against the core.

\begin{figure*}
  \centering
\includegraphics[height=8cm]{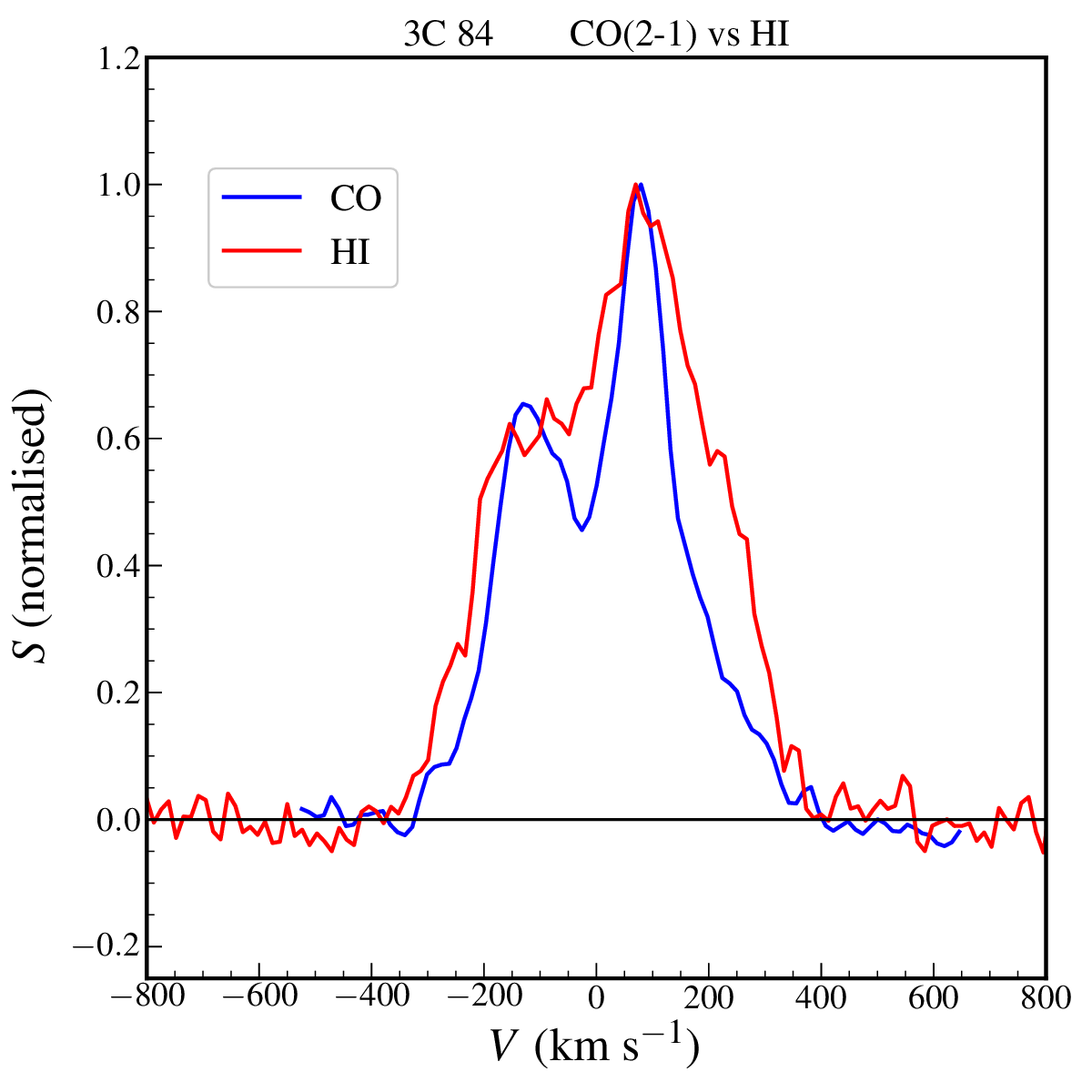}
\hskip0.5cm
\includegraphics[height=8cm]{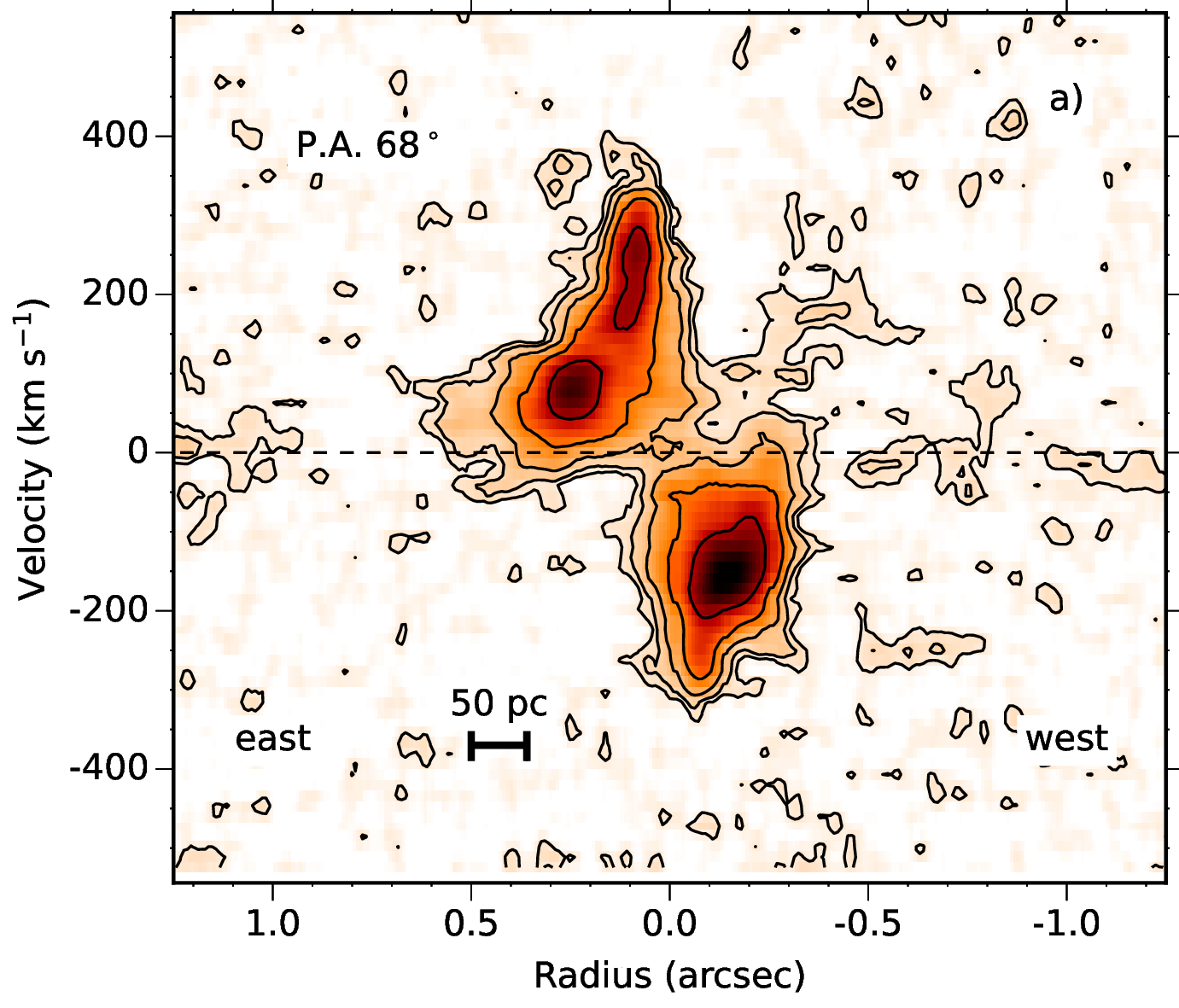}
 \caption{Comparison between the JVLA \HI\ absorption data and the kinematics of the molecular CND as observed with ALMA by \cite{Oosterloo23}.  {\it Left:} The JVLA \HI\ absorption profile (extracted from the central beam of the natural weighting cube and shown in red; inverted to facilitate comparison) and the  profile of the integrated \coTwo\ emission from the circum-nuclear molecular disc shown in blue \citep{Oosterloo23}. The \HI\ profile has been smoothed to the same velocity resolution as the \coTwo\ profile, and the profiles have been normalised to the peak. 
 {\it Right:} Position-velocity diagram of the circum-nuclear CO disc taken along the kinematic major axis (from \citealt{Oosterloo23}). The full range of velocities clearly only occurs for radii smaller than  50 pc. In both panels the  velocities are with respect to the systemic velocity of NGC 1275.
} 
\label{fig:ProfilesComparison}
\end{figure*}

\section{Discussion}
\label{sec:Discussion}

\subsection{Origin of the \HI\ absorption}
\label{sec:Origin}

The main observational result of this paper is that we clearly detect the  low-velocity \HI\ absorption against the central $\lta$1 arcsec region ($\lta$350 pc) of 3C~84 in the JVLA data, while there is no trace of this absorption in the higher-resolution VLBA observations. Together with the information from the WSRT observations, this allows us to determine the location of the absorbing gas and learn more about the properties of the circum-nuclear structures of 3C 84.

As mentioned above, the low-velocity absorber was also observed with the WSRT at a spatial resolution of \apx 15$''$. This scale is an order of magnitude larger than was achieved with the JVLA. A comparison of the WSRT and the JVLA spectra is shown in Fig.\ \ref{fig:Comparison}, and it is clear that the two spectra agree very well, both in terms of the depth of absorption an in  the shape of the absorption profile. This good agreement  implies that the gas giving rise to the low-velocity absorption is distributed on sub-arcsecond scales (i.e.\ $< 350$ pc, the size of the JVLA beam), and that the background continuum structure against which the absorption is seen is this large at most. 

Because the absorption comes from a small region, an obvious candidate for the continuum emission against which the absorption is seen would be the bright VLBA continuum structure, which has a size of about 50 mas (i.e.\  a few dozen parsecs), and which includes the core and the bright inner jet.  However, our VLBA data show no absorption against these components, even though the sensitivity of our observations is such  that we  would easily have detected this absorption if it originated there. This means that  the background continuum must be located away from the core (i.e.\ at radii larger than a few dozen parsec). It also implies  that the absorption seen in the VLA data is not due to a foreground cloud. An unrelated cloud like this would be expected to produce absorption against the 14 Jy core of 3C 84 in a similar way  as the high-velocity system.  

Another indication about the nature of the \HI\ structure causing the absorption is that the velocity range of the \HI\ profile is very similar to that of the molecular CND
as observed (in emission) in warm H$_2$ by \cite{Wilman05} and \cite{Scharwachter13} and in \coTwo\ by \cite{Nagai19} and \cite{Oosterloo23}.  The left panel in Fig.\ \ref{fig:ProfilesComparison}  shows a comparison between the integrated spectrum of the \coTwo\ emission  of the inner \apx 50 pc radius as obtained by Oosterloo et al., together with the \HI\ absorption profile (extracted from the central beam of the natural weighting cube). The two spectra clearly agree very well in terms of total width and of asymmetry. This suggests that the \HI\ absorption comes from atomic gas that is located in a structure similar to the molecular CND, which has a diameter of about 100 pc ($\sim 0.3$ arcsec) and extends in the direction roughly perpendicular to the jet.

\begin{figure}
  \centering
\includegraphics[width=\linewidth]{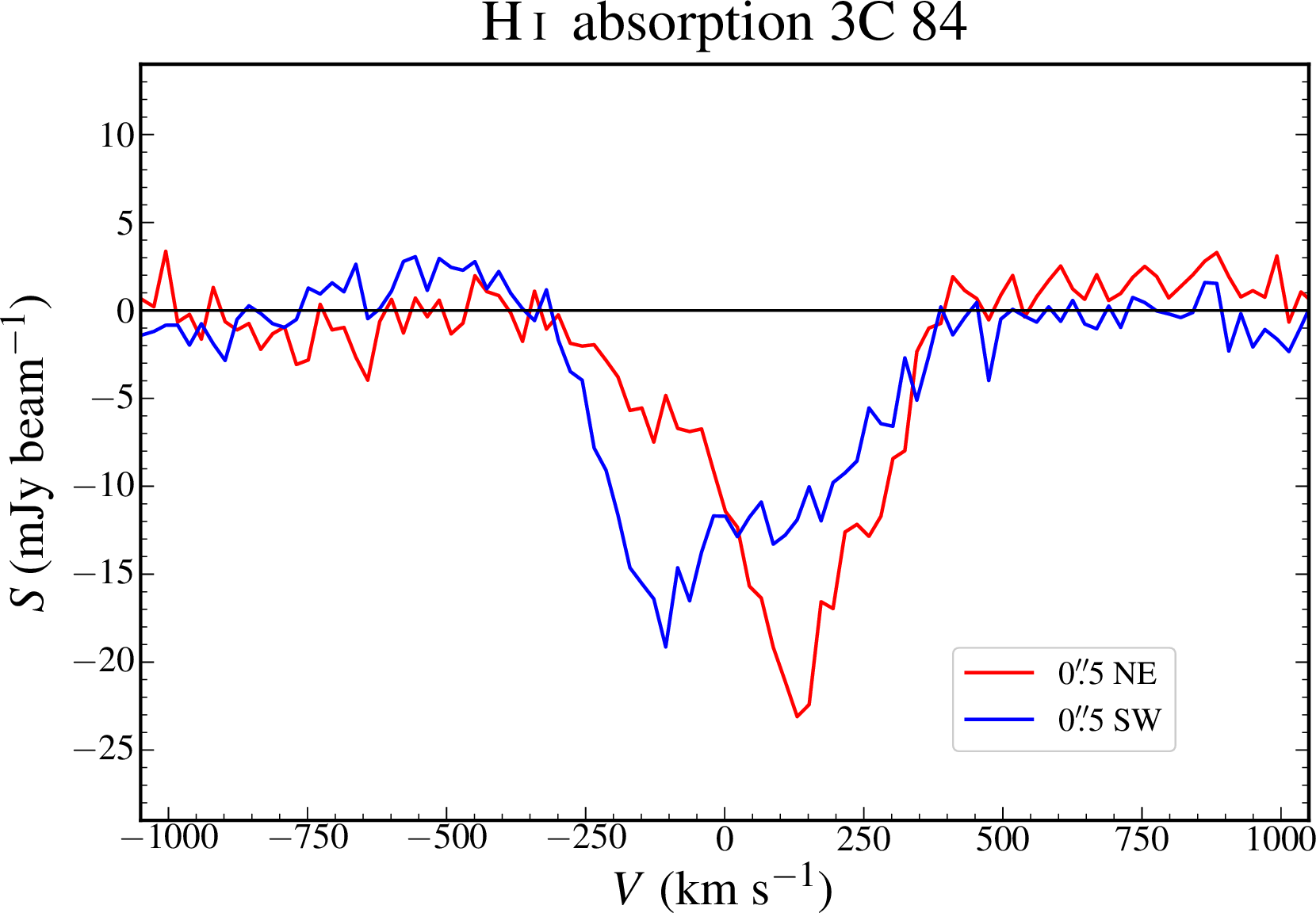}
 \caption{\HI\ spectra extracted from the JVLA robust $= -1$ data cube 0\farcs5 north-east and 0\farcs5 south-west of the peak continuum, showing the same rotation signature as the CND. For clarity, the spectra have been smoothed to a resolution of 21.6 \kms. Velocities are relative to systemic.}
\label{fig:ProfilesBothSides}
\end{figure}

To determine this, we extracted spectra from the JVLA data cube made with robust $= -1$, which has a resolution of  $1.10 \times 1.02$ arcsec, at two diametrically opposite locations, offset 0\farcs5 north-east and 0\farcs5 south-west from the peak continuum  of 3C~84  (Fig.\ \ref{fig:ProfilesBothSides}). Although the peak continuum is only very slightly resolved, we find that the peak of the \HI\ absorption shifts bluewards from the north-eastern part of the centre to the south-western part. This is exactly the  behaviour expected from  the kinematics of the CND, whose kinematical major axis has a position angle of 68$^\circ$ , and the NE part has receding velocities, as shown in the right panel  of Fig.\ \ref{fig:ProfilesComparison}. This is another strong indication that  the \HI\ gas seen in absorption  arises from the  molecular circum-nuclear gas disc and is probably located relatively close to the core because the highest velocities of the gas are observed there.

\begin{figure}
  \centering
\includegraphics[width=\linewidth]{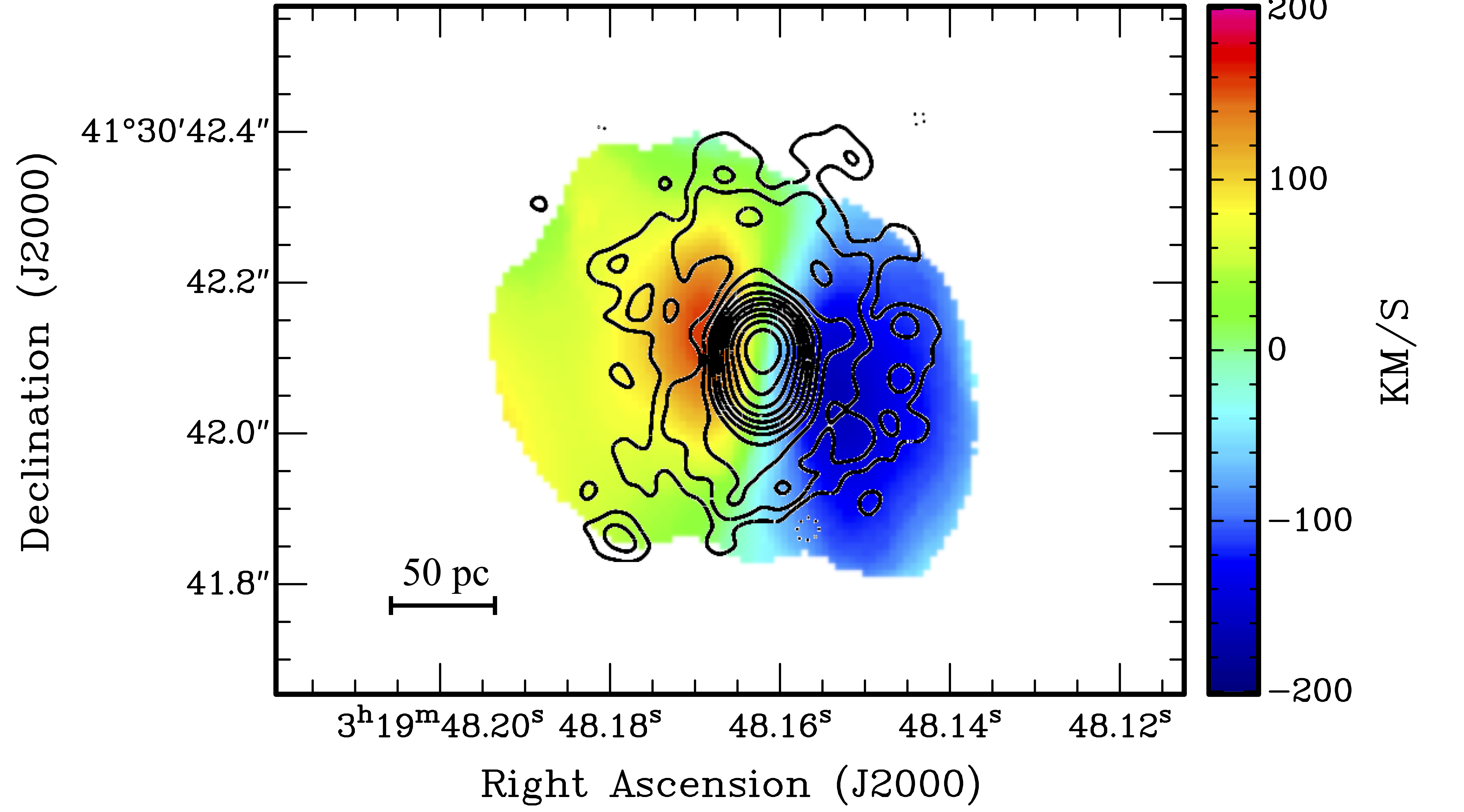}
 \caption{Overlay of the diffuse radio continuum emission at 330 MHz from the VLBA data of \cite{Silver98} and the velocity field of the central \coTwo\ disc from the reanalysed ALMA data presented in \cite{Oosterloo23}. }
\label{fig:continuum-CO}
\end{figure}

If the CND causes the low-velocity \HI\ absorption system, the radio continuum causing the absorption probably also spans similar spatial scales. Intriguingly, \cite{Silver98} used multi-wavelength VLBI observations and reported diffuse radio emission spanning \apx 100 pc that spatially overlapped with the circum-nuclear disc. \cite{Nagai21} reanalysed the  330~MHz VLBA data of  \cite{Silver98} and highlighted the spatial coincidence between this extended continuum emission and the molecular CND. We illustrate this in Fig.\ \ref{fig:continuum-CO}, where this extended continuum emission is shown on top of the velocity field of the molecular CND. \cite{Nagai19} argued that this strongly suggests that  the relativistic electrons emitting the radio continuum and the molecular gas disc are co-spatial, and that supernova explosions in the CND are at the origin of these electrons, and not the AGN. \cite{Silver98} reported that the radio luminosity of the CND agreed well with the infrared luminosity as seen in normal galaxies \citep{Condon92}. This indicates on-going star formation, which is further supported by the high density of the molecular gas. Based on the calculations and assumptions presented in \cite{Silver98}, a star formation rate (SFR) of 3 \msunyr\ is expected (see the discussion in \citealt{Nagai21}). This SFR is also capable of inducing turbulence in the CND, where \cite{Nagai19} and \cite{Oosterloo23} observed a relatively high velocity dispersion of \apx 25 \kms. All this strongly suggests that the \HI\ absorption occurs  against this radio continuum, which arises from the CND itself.

Finally, another line of investigation also supports the idea that the background radio continuum emission very likely does not arise from the radio core and jets: the low-velocity absorber has been observed at least four times over the last four decades \citep[][ and the present work]{Crane82, Sijbring93, Morganti20}. In this time span, the radio AGN has shown significant variability of over 10 Jy at milliarcsecond scales \citep[e.g.]{Taylor96, Silver98, Momjian02}. However, remarkably, the depth of the absorbed flux reported by all these studies has remained the same. Because we do not expect such dramatic changes (and on such short timescales) in the optical depth and column density of the screen producing the \HI\ absorption, this implies that the absorption does not occur against the variable radio component  because if it did, it should have changed in flux in step with the continuum.  \cite{Sijbring93} hypothesised that the absorption may arise only against the milliarcsecond-scale radio continuum, which is not variable. However, our VLBA observations clearly show that this is not the case either.

\subsection{Properties of the \slHI\ gas and the CND } 
\label{sec:Properties}

Although we determined the location of the gas producing the \HI\ absorption, we cannot derive the exact distribution and extent of the \HI\  on subkiloparsec scales. 
Furthermore, with the available data, it is difficult to comment on whether a vertical stratification of multiphase gas is present in the CND, as predicted by numerical simulations \citep{Wada16}. However, we can attempt to derive some of the properties of the CND from our observations.  Figure \ref{fig:ProfilesComparison}   presents a position-velocity diagram of the molecular CND, taken from the \coTwo\ data published by \cite{Oosterloo23} along the kinematic major axis of the CND (position angle 68$^\circ$). This shows the semi-Keplerian behaviour of the CND (see \citealt{Nagai19} and \citealt{Oosterloo23} for details), and in particular, that the full velocity range of the \coTwo\ only occurs in the inner part of the CND. This suggests  that at least part of the \HI\ absorption comes from this inner region. Assuming the gas is in Keplerian motion, we can convert  the highest detected velocities into the radius at which gas must be present. The largest deviations in the \HI\ profile from the systemic velocity are about 300 \kms\ , which gives a maximum deprojected velocity of 425 \kms\ when the inclination
of the CND is assumed to be about 45$^\circ$ \citep{Scharwachter13}. \cite{Bettoni03} estimated the mass of the SMBH to be $M_{\rm BH} = 10^{8.61}$ \msun. Based on this mass, this maximum velocity means that the smallest radius at which gas is present is about 20 pc.   \cite{Nagai19} estimated the sphere of influence of the SMBH in 3C~84 to be 60 pc. Thus, the \HI\ appears to reach this inner region despite the influence of the active SMBH.

That the broad low-velocity \HI\ absorption is not seen against the core and inner jet of 3C~84 but against  much fainter diffuse continuum emission has strong implications for estimating the column density of the \HI\ gas. 
\cite{Silver98} not only detected this diffuse emission at 330 MHz, but were also able to estimate its flux at 1.4~GHz  (and thus its spectral index). To do this, they tapered their  1.4-GHz VLBA data extremely, which allowed them to detect some  of the extended continuum emission at this frequency as well, and they derived an integrated flux of  80 mJy  at 1.4~GHz for the diffuse emission. If  the \HI\ absorption detected with the JVLA is seen against this diffuse CND emission, it would mean that the gas has a high optical depth of $\tau \sim  0.80$  and thus a high column density of  $4.3 \pm 0.2 \times 10^{22}$ cm$^{-2}$  , assuming $T_{\rm spin} = 100 $ K and  $c_{\rm f} = 0.5$. However, the conditions of \HI\ gas located so close to the AGN  might well be such that the spin temperature is at least 1000 K, leading to column densities above $10^{23}$ cm$^{-2}$. These column densities are quite similar to those observed for the warm and cold molecular gas (\citealt{Scharwachter13,Oosterloo23}).  

\cite{Nagai21} used the approach presented in \cite{Kawakatu20} to derive  the thickness of the CND. They concluded that it is expected to be thin, with a height of about 10 pc at a radius of 100 pc. 
We do not see \HI\ absorption against the central core and the beginning of the jet. This gives some weak constraints on  the thickness of the \HI\ distribution.  If the CND has a constant thickness, there has to be a central hole in the CND, otherwise, we would always see gas in absorption. We estimated the radius of this hole to be about 20 pc above. Inside this radius, the gas is likely ionised throughout. When we take the inclination of the CND to be about 45$^\circ$, this means that the CND cannot be thicker than the size of the central hole, otherwise, absorption would still be observed, implying a thickness of at most 20 pc. This may suggest the \HI\ is distributed in a thicker disc than the molecular gas, similar to what was found in the case of Circinus \citep{Izumi18} and is predicted by numerical simulations \citep{Wada16}. However, when we assume the CND to be flaring (i.e.\ a thickness proportional to the radius), our data cannot set  constraints on the thickness of the distribution of the \HI. 

\subsection{Similarity with Mrk~231 and overall relevance}

Our data show that the broad \HI\ absorption detected in 3C~84 originates from the CND of this AGN. It is particularly interesting to note that this finding bears a striking similarity to the case of Mrk~231 as presented by \cite{Carilli98}. Mrk 231 is one of the very few objects for which the multi-phase structure of the CND, including the continuum emission from its star formation, has been studied in detail.

In Mrk~231,  broad \HI\ absorption is also detected on arcsecond scales, but not on milliarcsecond scales. Based on this, \cite{Carilli98} suggested that the absorption comes from \HI\ in the circum-nuclear disc which has a size of about 60 pc, instead of being caused by clouds in front of the radio jet.  \cite{Taylor99} concluded that  diffuse synchrotron emission is present in Mrk 231, which is most likely related to  star formation activity in the CND. \cite{Carilli98}  were able to confirm their scenario by directly detecting the \HI\ absorption against the continuum emission from the star-forming CND by using the short baselines of  the VLBA (including the phased-up VLA). Unfortunately, our 3C~84 VLBA data do not have enough sensitivity on the shorter baselines to produce a deep enough image to trace such a disc at 1.4~GHz in 3C~ 84. The strength of the continuum, limiting the dynamic range reached, creates an additional complication, and dedicated observations are needed. 
Another similarity is that the optical depth of the absorption against the faint CND in Mrk 231 is also very high ($\tau = 0.17 \pm 0.02$), implying column densities of about $10^{23}$ cm$^{-2}$.  

 3C 84 and Mrk 231  are very different types of radio sources and  also live in very different  environments. Their CNDs appear to have very similar structures, however. This might suggest that these CNDs   are relatively common, but can only  be identified when a rich set of observations is available.

\subsection{Impact of the radio jet}

A number of radio continuum studies have suggested interaction between the radio plasma and gas clouds in the central region in order to explain the observed change in position angle of the radio jet (e.g.\ \citealt{Nagai17,Kino18} and references therein). This is particularly the case for the C3 component in the inner jet, which is located only a few milliarcseconds (i.e.\ a few parsec) from the SMBH. At the resolution of our observations, this region is included in our C component.
At this location, a hot spot was detected that changed position at the various epochs of the available observations. This could  indicate that the jet meanders in an inhomogeneous ambient medium, which results in the movement of the termination shock,  or that it is due to  time-variable ejection of jet flows \citep{Nagai17}.
In the scenario described above as obtained from our \HI\ observations, the lack of \HI\ absorption on  parsec scales indicates that no \HI\ is seen at the location of the jet and, most importantly, no kinematically disturbed cold gas is seen to result from gas interacting with the radio jet.  

Because the cold gas seems to be absent in the central region ($<20$ pc), it is interesting to study whether these interactions might instead be present in the ionised gas. A large amount of ionised gas in the central regions has been traced  by the detection of free-free absorption \citep{Walker00} and of emission lines. Interestingly, when emission lines are used ([FeII] and Br$\gamma$),  ionised gas is also seen to be distributed in a CND, which  includes gas characterised by a relatively high velocity dispersion \citep{Scharwachter13} and possibly also outflowing  gas resulting from the shocks \citep{Riffel20}. However, even in the latter case, the data do not show any coincidence between the possibly disturbed kinematics of the gas and the radio jet.  
We did not detect gas with disturbed kinematics at the core location of  \HI\ either, even though the upper limit to the optical depth of the gas against the core C region  is quite low, and lower than the typical values seen for the shallow and blueshifted \HI\ absorption tracing jet-cloud interactions as seen in some radio galaxies (e.g.\  \citealt{Morganti05,Mahony13}).

Interestingly,  another possible explanation could be that the observed change in position angle of the jet is instead due to  wobbling of the CND caused by  slight changes in the direction of the ongoing gas deposition from the surrounding filaments of molecular gas onto the disc. This would result in a small reorientation of the CND and hence of the jet (see the discussion in \citealt{Oosterloo23}). 

If the jet and the gas in the centre of 3C 84 do not interact directly, the type of feedback produced by 3C 84 is affected as well.  
It suggests that in this source, the jet has already opened a channel in the surrounding gaseous medium and has already expanded  beyond the gas-rich central part of the galaxy. Because of this, we do not observe gas outflows driven by the jet, and the effect of the radio jet seems to have  shifted to the "maintenance mode",  creating  bubbles in the circum-galactic medium (CGM) and preventing much of the cooling  of the  CGM on galaxy scales and larger (as also suggested by \citealt{Lim08}).
Because of the size of the radio jet, which in 3C~84 extends to several kiloparsec, the scenario is consistent with what is seen in other radio galaxies (e.g.\  PKS~0023--63 and 3C~305; \citealt{Morganti21,Morganti23}), as well as in low-luminosity radio sources (e.g.\ \citealt{Rosario19,Feruglio20,Cresci23}). Thus, the impact of the radio jets on producing outflows appears to be  limited to the first phase of their evolution, while at a later stage, their main impact is in the maintenance mode, affecting galaxy scales and beyond (see also \citealt{Oosterloo23}).

\section{Conclusions} 

We have presented a multi-resolution study of the \HI\ absorption in 3C~84. The combination of these data with information on the circum-nuclear structure seen in molecular gas at a similar resolution, as well as detailed images of the radio continuum, have allowed us to expand our understanding of this iconic radio galaxy. We concentrated on the \HI\ absorption centred on the systemic velocity of NGC~1275. Our data confirm that this absorption  originates from the inner 350 pc (the scale of the JVLA A-array beam) central region. 
On the other hand, the higher-resolution (a few pc) VLBA data showed that the absorbing gas is not located against the core or inner jet. 

The striking agreement we find between the velocities and shape of the \HI\ profile and those of the molecular gas of the CND (\citealt{Nagai19,Oosterloo23}) strongly suggests that the \HI\ absorption is associated with the CND.  This is further confirmed  by  using the JVLA data at the highest possible resolution to show that apart from the velocity range, the \HI\ absorption shows the same spatial kinematic signature as the CND.

These results tell us about the composition of the CND, and they now can form the basis for  more detailed studies  to investigate whether stratification is present in the different phases of the gas, as predicted by simulations \citep{Wada16}.

The star formation in this circum-nuclear structure appears to be the key not only to explain the   radio continuum emission, which provides the background for detecting \HI\ absorption, but also to explain the feeding of the AGN by inducing turbulence to the gas \citep{Nagai19}.

The similarity of the circum-nuclear \HI\ (and cold molecular) properties between  3C~84 and Mrk~231, combined with the fact that these two objects are quite different as radio sources and in terms of their environment, suggests that  CNDs with these properties may be relatively common. Observations with a wide dynamic range and high sensitivity are needed to explore this further.

We confirm the presence of \HI\ in 3C84 down to  \apx 20 pc from the SMBH through the comparison of the  velocities observed with respect to those of the molecular gas observed in emission. This  occurs despite possible effects of the AGN on the physical conditions of the gas. 
However, the non-detection of absorption against the core and inner jet suggests that the gas in the very inner region is mostly ionised.  Thus, high spatial resolution \HI\ absorption observations do not necessarily trace the \HI\ gas in the very inner regions, and it is essential to have a good knowledge of the radio continuum that may provide the background for the absorption:  sensitive observations with different spatial resolution are crucial for this. Furthermore, complementing the \HI\ data with information on the  molecular gas has been the key for the interpretation. 

We did not detect kinematically disturbed \HI\ gas despite the low optical depth reached by our observations. This result supports the picture in which the impact of jets may change as they evolve and expand farther into the ISM. Although new components are regularly ejected from the nucleus of 3C~84 that feed the jet, the fact that the overall jet is already evolved and many dozen kiloparsec in size  suggests that it has already opened a channel in the ISM and that its main impact is in a maintenance mode and not in producing gas outflows. This is consistent with the proposed scenario in which the jets, while travelling through the hot cluster gas, create cavities in the hot cluster gas, dump energy into the cluster gas, and quench to some extent the cooling flow of the hot ICM of the Perseus cluster (\citealt{Salome06,Lim08,Fabian11}). On the other hand, around the edges of these rising plasma bubbles, the gas is compressed, which induces cooling of this gas, leading to the formation of the filaments of warm and cold gas (see e.g. \citealt{Lim08}). These filaments can fuel the CND, which in turn is likely to cause the fuelling of the AGN \citep{Nagai19,Oosterloo23}. The results from our \HI\ observations are consistent with this scenario. 

Our study has highlighted the importance of combining the view of the line and the radio continuum on different scales (up to high spatial resolution) and of doing this for different phases of the gas to interpret \HI\ absorption studies of AGN.

\begin{acknowledgements}
CC acknowledges support via the ASTRON/JIVE summer student program. HN is supported by JSPS KAKENHI grants No. JP18K03709 and No. JP21H01137. SM thanks Junghwan Oh and Gabor Orosz for useful discussions on the VLBA data reduction. The National Radio Astronomy Observatory is a facility of the National Science Foundation operated under cooperative agreement by Associated Universities, Inc. 
The Westerbork Synthesis Radio Telescope was operated by ASTRON (Netherlands Institute for Radio Astronomy) with support from the Netherlands Foundation for Scientific Research (NWO).
This paper makes use of the ALMA data of ADS/JAO. ALMA\#2017.0.01257.S. 
ALMA is a partnership of ESO (representing its member states), NSF (USA) and NINS (Japan), together with NRC (Canada), MOST and ASIAA (Taiwan), and KASI (Republic of Korea), in cooperation with the Republic of Chile. The Joint ALMA Observatory is operated by ESO, AUI/NRAO and NAOJ.

\end{acknowledgements}

{}


\begin{appendix}

\section{Two high-velocity absorbers}

\begin{table*}[h]
\caption{Parameters of Gaussian fits to the newly discovered absorption system}
\centering
\begin{tabular}{ccccccccccc}
\hline\hline
component & $V_{\rm hel}$ & $\tau_{\rm max}$ & $\sigma$  & $\int \tau\,dv$ & $N_{\rm HI}$ \\
 & (\kms) & & (\kms) & (\kms) &(cm$^{-2}$)  \\
(1) & (2) & (3) & (4) & (5) &(6) \\ 
\hline
1 &$7934.5 \pm 0.4$  & $0.0010 \pm 0.0001$ & $2.7  \pm 0.4$ &
  $0.0070 \pm 0.0013$ & $1.3 \pm  0.2 \times 10^{18}$\\
2 & $7917.7 \pm  0.5$ &$0.0008 \pm 0.0001$ & $1.7   \pm  0.3$ &
$0.0033  \pm  0.0009$ & $5.9 \pm 1.6 \times 10^{17}$\\
\hline
 \hline
\end{tabular}
\begin{tablenotes}
\item The columns list (1) the component, (2) the central velocity, (3) 
the peak optical depth, (4) the velocity dispersion not corrected for instrumental broadening, (5) the integrated optical depth, and (6) the \HI\ column density assuming $T_{\rm spin} = 100$ K.
\end{tablenotes}
\label{tab:hvsTab}
\end{table*}

\begin{figure}[h]
  \centering
\includegraphics[width=\linewidth]{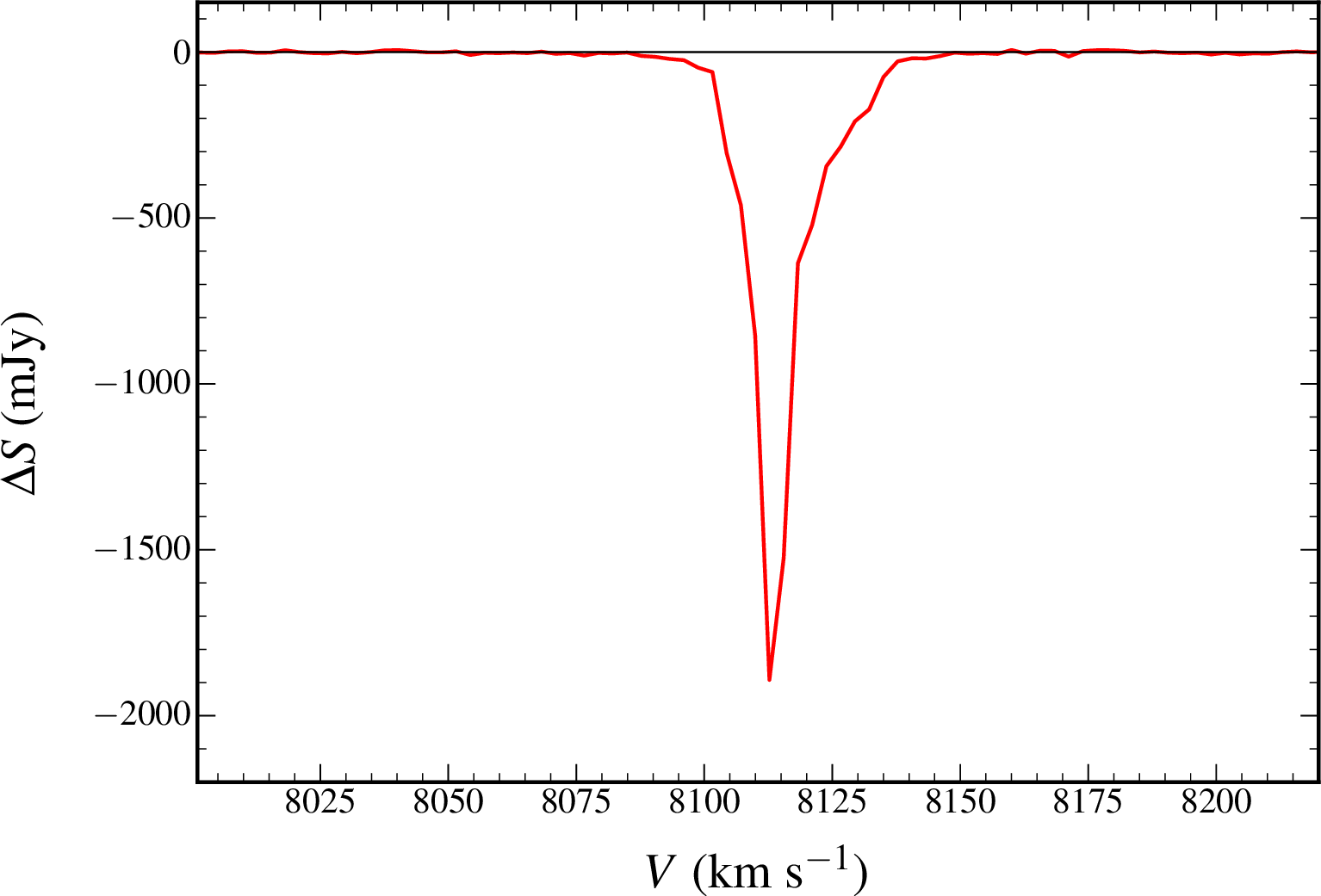}
 \caption{Spectrum of the high-velocity absorption system redshifted 2850 \kms\ with respect to the systemic velocity of NGC 1275. It was earlier studied by \cite{Momjian02}.}
\label{fig:oldHigh}
\end{figure}

In addition to the broad \HI\ absorption detected at the systemic velocity of NGC 1275, two other absorption systems were detected in the JVLA data. One of these is a deep, narrow absorption line  at $V_{\rm hel}$ = 8115 \kms\ (Fig.\ \ref{fig:oldHigh}) that was discovered by \cite{DeYoung73} and   was studied at high spatial and velocity resolution with the VLBA by \cite{Momjian02}. This absorption is  associated with a foreground cluster galaxy, commonly referred to as the high-velocity system (HVS), that is falling towards NGC 1275, but is still \apx 100 kpc distant from the AGN host galaxy. The HVS was first identified by \cite{Minkowski55} and causes most of the dust lines seen in optical images of NGC 1275. It is also detected in optical emission lines \citep[e.g.][]{Yu15} and in absorption in X-rays \citep[e.g.][]{Sanders07}. 

For this absorption, we measure an integrated optical depth of $\int \tau\ dv = 1.63 \pm 0.07$ \kms\ 
, which gives a column density of $2.9 \pm 0.1 \times 10^{20}$ cm$^{-2}$ assuming $T_{\rm spin} = 100 $ K. This is consistent with the results of \cite{Momjian02}.

\begin{figure}[h]
  \centering
\includegraphics[width=\linewidth]{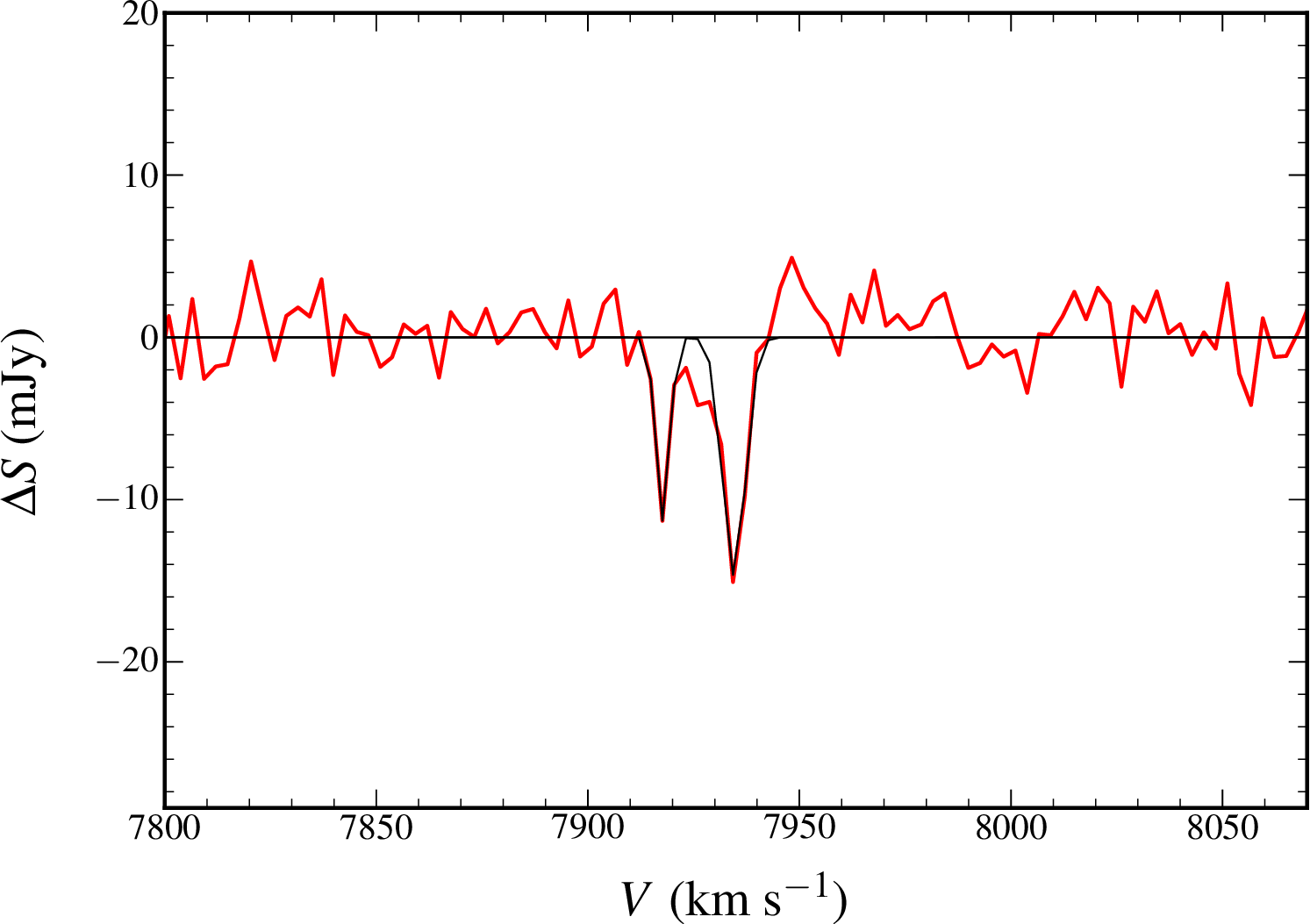}
 \caption{Spectrum of the newly discovered absorption system redshifted by 2660 \kms\ with respect to the systemic velocity of NGC 1275. The two Gaussian components fitted to the profile are indicated in black. Their parameters are listed in   Table \ref{tab:hvsTab}.}
\label{fig:newComp}
\end{figure}

In addition to this previously known HVS absorption system, we discovered a much fainter absorption component redshifted by 2660 \kms\ with respect to the systemic velocity of NGC 1275 (Fig.\ \ref{fig:newComp}), but slightly blueshifted with respect to the HVS. This component is also likely  associated with the infalling galaxy because ionised gas with  velocities blueshifted up to 800 \kms\ with respect to those of the main body of the HVS has been found in optical data \citep{Yu15}. This gas is likely stripped from the HVS by the cluster medium, and this absorption is likely part of this gas.

This newly discovered absorption system appears to consist of two components. We fitted Gaussians to these two components, the parameters of which are given in Table \ref{tab:hvsTab}. The column densities of the two components are   $5.9 \pm 1.6 \times 10^{17}$ cm$^{-2}$ and $1.3 \pm  0.2 \times 10^{18}$ cm$^{-2}$ (assuming $T_{\rm spin} = 100 $ K).

\end{appendix}

\end{document}